\documentclass[aps,superscriptaddress,twocolumn,twoside,floatfix,pra,nofootinbib,a4paper]{revtex4-2}
\usepackage{enumerate,appendix}
\usepackage{amsmath, amsthm, commath}
\usepackage{color,calc,graphicx}
\usepackage[dvipsnames,svgnames,table,cmyk]{xcolor}
\usepackage[colorlinks]{hyperref}
\hypersetup{
	colorlinks = true,
	urlcolor = {blue},
	citecolor = {magenta},
	linkcolor= {blue}
}

\usepackage{graphicx}
\usepackage[normalem]{ulem}
\usepackage{amsmath}
\usepackage{latexsym}
\usepackage{bbm}
\usepackage{physics}
\usepackage{comment}
\usepackage[charter,cal=cmcal,sfscaled=false]{mathdesign}
\usepackage{booktabs}
\usepackage{multirow}
\usepackage{subfigure}
\usepackage{dcolumn}
\usepackage{diagbox}
\usepackage{array}
\usepackage{makecell}
\usepackage{threeparttable}
\hypersetup{
	colorlinks = true,
	urlcolor = {blue},
	citecolor = {magenta},
	linkcolor = {blue},
	breaklinks = true
}

\newtheorem{corollary}{Corollary}

\newtheorem{proposition}{Proposition}
\newtheorem*{result}{Main Result}
\theoremstyle{remark}
\newtheorem{remark}{Remark}

\begin{document}
\title{Unexpected consequences of Post-Quantum theories in the graph-theoretical approach to correlations}

\author{José Nogueira}
\email{josenogueira.castro@gmail.com}
\affiliation{Instituto de Física Gleb Wataghin, Universidade Estadual de Campinas, CEP 13083-859 Campinas, Brazil}

\author{Carlos Vieira}
\affiliation{Instituto de Matemática, Estatística e Computação Científica, Universidade Estadual de Campinas, CEP 13083-859, Campinas, Brazil}

\author{Marcelo {Terra Cunha}}
%\email{terra@email.com}
\affiliation{Instituto de Matemática, Estatística e Computação Científica, Universidade Estadual de Campinas, CEP 13083-859, Campinas, Brazil}

\begin{abstract}
This work explores the implications of the Exclusivity Principle (EP) in the context of quantum and post-quantum correlations.
We first establish a key technical result demonstrating that given the set of correlations for a complementary experiment, the EP restricts the maximum set of correlations for the original experiment to the anti-blocking set. 
Based on it, we can prove our central result: if all quantum behaviors are accessible in Nature, the EP guarantees that no post-quantum behaviors can be realized. 
This can be seen as a generalization of the result of [Phys. Rev. A 89, 030101(R)], to a wider range of scenarios.
It also provides novel insights into the structure of quantum correlations and their limitations. 
\end{abstract}

\maketitle

\section{Introduction}
Quantum Theory (QT) is one of the most successful scientific theories ever created by humankind; in that its predictions match with great accuracy empirical data. One paramount example is the loophole-free confirmation of the existence of quantum Bell-nonlocality \cite{Giustina2015, Hensen2015,Shalm2015,Hensen2016,Li2018}. Nonetheless, it remains a challenge to formulate it as a consequence of physically justifiable principles (of Nature), similar to what occurs with Einstein's Special Relativity. Several distinct approaches have been proposed to address this problem. For example, certain approaches adopt an axiomatic framework, beginning with a more physically intuitive set of axioms from which one aims to derive the standard and abstract postulates of quantum theory (the Hilbert space structure, density matrices and measurement operators acting on it, and Born’s rule) \cite{hardy2001quantumtheoryreasonableaxioms,hardy2011reformulatingreconstructingquantumtheory, Masanes2011, Chiribella2010, Chiribella2011}. Another program is the one started by Popescu and Rohrlich's seminal work in trying to answer the question: are quantum correlations singled out by the non-signaling principle \cite{Popescu1994}? The answer was found to be negative, as there exist correlations that satisfy the non-signaling condition but cannot be reproduced within the framework of quantum theory, with Popescu-Rohrlich boxes being the most prominent example \cite{Brunner2014}. Although this approach does not fully explain quantum correlations, it raises the broader question: What is the fundamental physical principle that explains quantum correlations\footnote{An equivalent way of phrasing this question is: given that there exist correlations satisfying relativity, but more Bell-non-local than the quantum ones, what in Nature prohibits their existence? Or even, why is Nature not more non-local? What limits Nature's non-locality? All things considered, this program's idea is to study the set of quantum correlations in Bell scenarios from the outside, investigating physically motivated device-independent restrictions strong enough to rule out, at best, all post-quantum behaviors.}? Several principles to address this question, named as \textit{the problem of quantum correlations}, have since then been proposed: Non-Triviality of Communication Complexity \cite{Brassard2006}, Macroscopically Locality \cite{Navascus2009}, Local Orthogonality \cite{Fritz2013} and Information Causality \cite{Pawowski2009,Pollyceno2023} being examples of such principles.
All those trials focus on the problem of distant correlations, with the bipartite Bell scenario used as a playground for initial discussions, which could be further expanded to multipartite scenarios.
In this work we have a special interest in the so-called Exclusivity Principle (EP) \cite{Cabello2013a, cabello2013proposedexperimentexcludehigherthanquantum, nawareg2013bounding, Cabello2013b, Yan2013, Amaral2014, Cabello2014, Cabello2015, jia2022exclusivity, Cabello2019, Cabello2019b}, which avoids the necessity of parts, dealing with the problem of quantum correlations in Kochen-Specker (KS) contextuality scenarios \cite{Budroni2022}.

In 2010, Cabello, Severini, and Winter (CSW) presented an interesting connection between graph-theoretic concepts and KS-contextuality \cite{https://doi.org/10.48550/arxiv.1010.2163, Cabello2014c}. This framework starts with a set of measurement events $e_1, \ldots, e_n$. 
Two events are considered exclusive if they correspond to different outcomes of the same measurement\footnote{A measurement event $e_i$ is to be understood as the performance of a given contextuality scenario, meaning $e_i = a, b, ..., c|x, y, ..., z$; in words, obtaining outcomes $a, b, ..., c$ conditioned, respectively, on the chosen measurement context $x, y, ..., z$. Two events $e = a, b, ..., c|x, y, ..., z$ and $e^{\prime} = a^{\prime}, b^{\prime}, ..., c^{\prime}|x^{\prime}, y^{\prime}, ..., z^{\prime}$ are said to be exclusive if $(x = x^{\prime} \land a \ne a^{\prime}) \lor (y = y^{\prime} \land b \ne b^{\prime}) \lor ... \lor (z = z^{\prime} \land c \ne c^{\prime})$.}. We associate each event $e_i$ with a vertex $i$, and we say that two vertices $i$ and $j$ are adjacent, denoted as $i \sim j$, if the corresponding measurement events $e_i$ and $e_j$ are exclusive. Based on these vertices and adjacency relations, we construct a graph $G$, which we refer to as the exclusivity graph.

For a given exclusivity graph $G$, we consider theories that assign probabilities to the measurement events corresponding to its vertices. 
A behavior associated with $G$ is represented by a vector $(p(e_1), \ldots, p(e_n))$, where $p(e_i)$ represents the probability of event $e_i$. Since any two exclusive events correspond to different outcomes of some measurement, the behavior of a graph must also satisfy the relation $p(e_i) + p(e_j) \le 1$, whenever $i \sim j$ \cite{amaral_graph_2018}.

The set of behaviors is determined by the exclusivity constraints defined by the graph $G$ and by the underlying physical theory used to describe the system. 
For instance, a behavior $(p(e_1), \ldots, p(e_n))$ is called deterministic noncontextual if it satisfies $p(e_i) \in \{0,1\}$ and $p(e_i) + p(e_j) \leq 1$ for all pairs $i \sim j$ \cite{bharti_robust_2019}. 
The set of all noncontextual (or classical) behaviors, denoted as $\mathrm{NC(G)}$, is then obtained by making probabilistic mixtures of such deterministic behaviors, a construction known as the convex hull of those deterministic noncontextual behaviors\footnote{In graph-theoretical terms, this set is known as $\mathrm{STAB}(G)$ \cite{Cabello2014c, Knuth1994}.}. 
Behaviors that do not belong to $\mathrm{NC(G)}$ are considered contextual.

Second, one behavior is said to be quantum if there exists a quantum state $\rho$ and a set of projectors ${\Pi_i}$ acting on a Hilbert space $\mathcal{H}$ such that $p(e_i) = \tr{\rho \Pi_i}$ and $\Pi_i \Pi_j = 0$ whenever $i \sim j$ \cite{amaral_graph_2018}. The set of quantum behaviors\footnote{In graph-theoretical terms, this set is referred to as $\mathrm{TH}(G)$ \cite{Cabello2014c, Knuth1994}.} is denoted by $Q(G)$.
In quantum theory, the pairwise joint measurability of a set of projective measurements implies the joint measurability of the entire set  \cite{specker_logik_1960}. This property no longer holds for POVMs; in this case, one can construct examples that are pairwise jointly measurable but not globally jointly measurable \cite{liang_speckers_2011, kunjwal_quantum_2014}.
For quantum behaviors, if a collection of measurement events ${e_k}$ is pairwise exclusive, the constraint $\sum_k p(e_k) \le 1$ must hold\footnote{Indeed, following the definition of a quantum behavior of an exclusivity graph, a set of pairwise exclusive events $\{e_k\}$ is associated with a set of orthogonal projectors $\{\Pi_k\}$; therefore, $\sum_k \Pi_k \le \mathbb{1}$ and $\sum_k p(e_k) = \sum_k \mathrm{Tr}(\rho \Pi_k) \le 1$ \cite{amaral_graph_2018}.}. The \textit{Exclusivity Principle} (EP) extends this property as a universal hypothesis, proposing that this constraint should be satisfied in any physical theory. 
In other words, the EP asserts that for any set of pairwise exclusive measurement events ${e_k}$—that is, events that can be associated to mutually exclusive outcomes of a single mother measurement—the sum of their probabilities must satisfy $\sum_k p(e_k) \leq 1$. 
We define EP$_{1}$-behaviors as those that adhere to the Exclusivity Principle. The set of all such behaviors is referred to as the (single-copy) EP set\footnote{In graph-theoretical terms, this set is referred to as $\mathrm{QSTAB}(G)$ \cite{Cabello2014c, Knuth1994}.}, denoted by $E_1(G)$.

Cabello, Severini, and Winter \cite{https://doi.org/10.48550/arxiv.1010.2163, Cabello2014c} showed that given any noncontextuality (NC) expression appropriately written as a positive linear combination of probabilities of measurement events, $S_{\omega} = \sum_{i} \omega_{i} p(e_{i})$, its corresponding classical (or noncontextual), quantum, and (single-copy) exclusivity bounds match three characteristic numbers of the associated weight-exclusivity graph, a weighted-graph created from the exclusivity graph associated to the events $e_1, \ldots, e_n$, and the weight-vector $\omega = (\omega_1, \ldots, \omega_n)$. 
Specifically, this relationship is expressed as 
\begin{equation}
S_{\omega} \le^{\mathrm{NC}} \alpha(G, \mathbf{\omega}) \le^{\mathrm{Q}} \vartheta(G, \mathbf{\omega}) \le^{\mathrm{E}_{1}} \alpha^{*}(G, \mathbf{\omega}),
\end{equation}
in other words, $\mathrm{max}_{\mathrm{NC(G)}} S_{\mathbf{\omega}} = \alpha(G, \mathbf{\omega})$, $\mathrm{max}_{\mathrm{Q}(G)} S_{\mathbf{\omega}} =\vartheta(G, \mathbf{\omega})$ and $\mathrm{max}_{\mathrm{E}_{1}(G)} S_{\mathbf{\omega}} = \alpha^{*}(G, \mathbf{\omega})$, where $\alpha(G, \mathbf{\omega})$ is the independence number, $\vartheta(G, \mathbf{\omega})$ is its Lovász number, and $\alpha^{*}(G, \mathbf{\omega})$ is its fractional packing number, being all these three numbers well-studied objects in graph theory (and combinatorial optimization) \cite{lovasz_shannon_1979, grotschel_relaxations_1986, grotschel_geometric_1993, Knuth1994, amaral_graph_2018}. 

Now, since the quantum bound of any NC inequality corresponds to the Lovász number of its associated exclusivity graph $(G, \mathbf{\omega})$, and $\vartheta(G, \mathbf{\omega})$ is obtained from the set of quantum correlations of $G$, $\mathrm{Q(G)}$, if a physical principle is able to single out the latter (or, equivalently, justify why post-quantum graph behaviors are not allowed), then such principle corresponds to an answer of why is Nature not more contextual. This line of reasoning was first proposed in 2013 by Cabello \cite{Cabello2013a}, bringing the problem of quantum correlations to the CSW approach.

By considering two copies of the pentagon graph (the exclusivity graph associated with the KCBS inequality, known for being the simplest NC inequality \cite{Klyachko2008}), Cabello \cite{Cabello2013a} was able to show that the maximum value allowed by EP to it is $\sqrt{5}$, which also equals its maximum quantum value. Shortly afterward, Yan proposed \cite{Yan2013} a more general construction, involving the consideration of any given experiment and what is called its respective complementary experiment. This approach proved highly effective for applying the EP and enabled the extension of previous results. 
All results to be recalled in this paper, as well as the original ones to be shown, are derived from such underlying construction.
Let us give it some special attention.

Let $\{e_i\}$ be a set of $n$ measurement events with exclusivity graph $G$ and $\{e_{j}^{\prime}\}$ a set of also $n$ measurement events but with exclusivity graph $\bar{G}$, where $\bar{G}$ is the complementary graph\footnote{Complementary graph $\bar{G}$ is defined with the same vertex set as $G$, \textit{i.e.}, $V(\bar{G}) = V(G)$, but its edge set is defined by relation $ij \in E(\bar{G}) \iff ij \notin E(G)$.} of $G$. 
One then defines a set of composite events $\{f_{ij} = e_i \land e_{j}^{\prime}\}_{ij}$ ($f_{ij}$ occurs iff $e_i$ and $e_{j}^{\prime}$ occurs). The exclusivity graph encoding the exclusivity structure of events $f_{ij}$ is given by the disjunctive product\footnote{Given graphs $G = (V(G), E(G))$ and $H = (V(H), E(H))$, $G * H$ is defined as: $V(G * H) = V(G) \times V(H)$; edge $(i, j^{\prime})(k, l^{\prime}) \in E(G * H)$ iff $ik \in E(G)$ or $j^{\prime}l^{\prime} \in E(H)$.} of $G$ and $\bar{G}$, namely $G*\bar{G}$. For any $G$, $\{f_{ii}\}_{i}$ corresponds to a subset of events in $G*\bar{G}$ consisting of only pairwise exclusive events. Consequently, the EP implies
\begin{equation}\label{I}
\sum_i p(f_{ii}) = \sum_i p(e_{i}, e_{i}^{\prime}) \le^{\mathrm{EP}} 1.
\end{equation}
At this point, it is further assumed that the experiments $G$ and $\bar{G}$ are completely independent, allowing the joint probabilities of local events to factorize as $p(e_{i}, e_{i}^{\prime}) = p(e_{i}) p(e_{i}^{\prime})$. Surprisingly, even under this assumption of independence, one experiment can impose constraints on the other through the EP inequality, Ineq.~\eqref{I}. 
Such a paradigm is hereafter referred to as Yan's construction. 
By making assumptions about the available behavior set of the complementary experiment and applying the resulting EP inequality, several significant results were obtained.

Yan first demonstrated \cite{Yan2013} that, given $\mathrm{Q(\bar{G})}$, EP singles out the maximum quantum value of inequality $S = \sum_i p_i$ for experiment $G$. However, it is worth mentioning that there are infinitely many other inequalities\footnote{Specifically, any inequality $S_{\omega}$ with weights differing from $1$ for each measurement event.}, which are based on the exclusivity structure of experiment $G$, for which Yan's result says nothing about. 
In 2014, Amaral, Terra, and Cabello (ATC) established a stronger result \cite{Amaral2014}: given $\mathrm{Q(\bar{G})}$, EP singles out $\mathrm{Q(G)}$. 
This effectively resolves the limitation of the previous finding. 
Since EP picks out the entire quantum behavior set, it follows that for any chosen inequality $S_{\omega}$, EP also identifies its quantum maximum. Furthermore, ATC demonstrated that when the experiment consists of self-complementary graphs\footnote{A graph $G$ is self-complementary if it is isomorphic to its complement $\bar{G}$; an example is the KCBS graph.}, EP forbids Post-Quantum sets\footnote{Where by Post-Quantum sets we mean sets strictly larger than $\mathrm{Q(G)}$.}. A logical next step would be to ask whether such a result holds even without the self-complementarity assumption, \textit{i.e.}, to show that EP forbids sets larger than the quantum for any given experiment. In 2019, Cabello demonstrated \cite{Cabello2019,Cabello2019b} through a series of intricate arguments — for instance, considering an infinite number of copies of an experiment --- that EP forbids sets larger than $\mathrm{Q(G)}$ for self-complementary graphs. A second key result in the same paper provides a systematic procedure for constructing a self-complementary graph $H(G)$ from any given initial graph $G$. By combining this graph construction technique with his first result and using some additional involved results from graph theory, Cabello was ultimately able to show that EP singles out $\mathrm{Q(G)}$ for any experiment. 
In some sense, the 2019 work solves the quantum correlations problem within the CSW approach. 
However, it does so by employing complex arguments, including constructions that extend far beyond those proposed by Yan. 

In this work,  we use Yan's construction together with the notion of anti-blocking theory \cite{Fulkerson1972} to show that correlation sets strictly larger than $\mathrm{Q(G)}$ are ruled out by the EP. 

\section{Results}
We begin by presenting the core technical contribution of this work in Proposition \ref{prop1}, where we establish that if the set of correlations for the complementary experiment is given by $\mathrm{X(\bar{G})}$, then the EP constrains the maximum set of correlations for the original experiment to the anti-blocking set of $\mathrm{X(\bar{G})}$. A direct consequence of this proposition, combined with a foundational graph-theoretical result \cite{Knuth1994}, leads to our first result: assuming the set of correlations for the complementary experiment is the $\mathrm{E_{1}(\bar{G})}$, the EP restricts the set of correlations for the experiment to the classical set (Corollary \ref{cor1}). Lastly, we present the central result of this paper: assuming that Nature is capable of realizing every behavior allowed by quantum theory and that the EP holds, it follows that post-quantum correlations are forbidden. This can be seen as a generalization of ATC’s second result to any exclusivity graph. In doing so, our work addresses a methodological convolutedness in the literature, where the 2019 study deviated from the established trend of utilizing Yan’s construction in favor of a significantly more complex approach. By employing Yan’s framework directly, we pave this road and demonstrate that results similar to those in the 2019 work can be obtained through a much simpler methodology, offering additional insights into its findings.

Let us start with a remark. 
\begin{remark}
Any behavior $p^{\prime}$ achievable in the complementary experiment $\bar{G}$, imposes a constraint on the independent experiment $G$.
\end{remark}
Indeed, under Yan's construction, EP inequality \eqref{I} becomes
\begin{equation}\label{II}
\sum_{i} p(e_{i}) p(e_{i}^{\prime}) \le 1.
\end{equation}
The above inequality can be seen as a half-space separation over the set of behaviors for the experiment $G$, where the probabilities $p(e_i^{\prime})$ are seen as the coefficients of this inequality. Moreover, we can progressively consider broader sets of achievable behaviors for the complementary experiment, and subsequently investigate the behavior sets of the experiment as defined by all resulting EP constraints of Ineq.~\eqref{II}.

\begin{proposition}\label{prop1}
Given that the correlation set describing the complementary experiment is $\mathrm{X(\bar{G})}$, the largest correlation set allowed by the EP for the experiment is $\mathrm{Y(G)} = \mathrm{abl}\: \mathrm{X(\bar{G})}$.
\begin{proof}
On the one hand, $\mathrm{Y(G)}$ is defined by all EP inequalities in form of Ineq.~\eqref{II}, which arise when considering all behaviors of the complementary experiment $\bar{G}$; this can be expressed mathematically as follows:
\begin{equation}\label{defY_1}
\mathrm{Y(G)} = \{p \ge 0 \:|\: \sum_i p_i p_{i}^{\prime} \le 1 \: \forall \: p^{\prime} \in \mathrm{X(\bar{G})}\}. 
\end{equation}
On the other hand, for any set of non-negative vectors $A$, its \textit{anti-blocking} associated set, denoted $\mathrm{abl}A$, is given by
\begin{equation}\label{defablA}
\mathrm{abl}A = \{b \ge 0 \:|\: \sum_i a_i b_i \le 1 \: \forall \: a \in A\}.
\end{equation}
Straightforward comparison between Eq.~\eqref{defY_1} and Eq.~\eqref{defablA} reveals that, under Yan's construction, EP constraint Ineq.~\eqref{II} matches exactly the defining property of an \textit{anti-blocking} set. Therefore, $\mathrm{Y(G)} = \mathrm{abl}\: \mathrm{X(\bar{G})}$.
\end{proof}
\end{proposition}

Although Proposition \ref{prop1} holds for all sets $\mathrm{X(\bar{G})}$, we will focus on the most relevant cases. 
These range from the smallest possible set to describe $\bar{G}$, corresponding to classical (or noncontextual) behaviors, $\mathrm{NC(\bar{G})}$, to the largest possible set consistent with the EP, corresponding to correlations that are only required to satisfy single-copy exclusivities, denoted $\mathrm{E_{1}}(\bar{G})$. 

In the following, we investigate the restrictive power of EP in selecting correlation sets for a given experiment, by assuming specific sets for the complementary experiment, including the limiting cases of classical and $E_1$ correlations, as well as the quantum one. 

\begin{corollary}\label{cor1}
Given $\mathrm{E_{1}(\bar{G})}$, EP singles out $\mathrm{NC(G)}$.  
Conversely, given $\mathrm{NC(\bar{G})}$, EP singles out $\mathrm{E_{1}(G)}$.
\begin{proof}
By Proposition \ref{prop1}, the largest correlation set allowed by the EP to the experiment is $\mathrm{Y(G)} = \mathrm{abl \:}\mathrm{E_1(\bar{G})}$. However, as shown in Ref.~\cite{Knuth1994}, for any experiment, $E_1$ and classical correlation sets are connected via the anti-blocking operation, specifically $\mathrm{E_1(\bar{G})} = \mathrm{abl \:}\mathrm{NC(G)}$. Consequently, it follows that $\mathrm{abl\:E_1(\bar{G})} = \mathrm{abl\:abl\:NC(G)}$. Since for any convex corner\footnote{A set of vectors $V$ is a convex corner if it is nonempty, closed, convex, nonnegative and $0 \le \vec{u} \le \vec{v}$ and $\vec{v} \in V \Rightarrow \vec{u} \in V$; see section $30$ of \cite{Knuth1994}.} $D$, we know that $\mathrm{abl \: abl \:}D = D$ and both $\mathrm{E_1(\bar{G})}$ and $\mathrm{NC(G)}$ are convex corners \cite{Knuth1994}, it results that $\mathrm{Y(G)} = \mathrm{abl\:E_1(\bar{G})} = \mathrm{NC(G)}$. As for the second part, the distinction between the label of an experiment and its complement is entirely arbitrary. Therefore, by exchanging $G$ with $\bar{G}$, we obtain $\mathrm{E_1(G)} = \mathrm{abl \: NC(\bar{G})}$. 
\end{proof}
\end{corollary}

The first part of Corollary \ref{cor1} states that if EP holds and there exists a complementary experiment $\bar{G}$ where it is provable that all EP$_{1}$-behaviors are accessible\footnote{The case of interest being an imperfect graph; for perfect graphs $\mathrm{NC(\bar{G}) = Q(\bar{G}) = E_{1}(\bar{G})}$.}, then to the experiment $G$, only noncontextual behaviors are allowed. 
As for the second part, if the EP is the only restriction and only classical behaviors are accessible to $\bar{G}$, then to the experiment $G$, all {EP}$_{1}$-behaviors are allowed. Notably, the larger the correlation set assumed for the complementary experiment, the stronger EP becomes in excluding behaviors from the experiment. Conversely, the smaller the correlation set assumed for the complementary experiment, the weaker EP becomes in excluding behaviors from the experiment. This mechanism is precisely what underpins the proof of our main result. Before showing it, we remark that Proposition \ref{prop1} recovers result $1$ due to ATC in Ref.~\cite{Amaral2014}.

\begin{remark}{(Result 1 of Ref.~\cite{Amaral2014})}\label{cor2} 
Given $\mathrm{Q(\bar{G})}$, EP singles out $\mathrm{Q(G)}$. 
\begin{proof}
By Proposition \ref{prop1}, the largest correlation set allowed by EP to the experiment $\mathrm{G}$ is $\mathrm{Y(G)} = \mathrm{abl \:}\mathrm{Q(\bar{G})}$. 
As shown in Ref.~\cite{Knuth1994}, for any experiment $\mathrm{G}$ the equality $\mathrm{Q(\bar{G})} = \mathrm{abl \: Q(G)}$ holds. Consequently, $\mathrm{abl \: Q(\bar{G})} = \mathrm{abl \: abl \: Q(G)} = \mathrm{Q(G)}$, since both $\mathrm{Q(G)}$ and $\mathrm{Q(\bar{G})}$ are convex corners. Thus, it follows that $\mathrm{Y(G)} = \mathrm{abl \: Q(\bar{G})} = \mathrm{Q(G)}$.
\end{proof}
\end{remark}
Moreover, we can see Proposition \ref{prop1} as a generalization of Result $1$ of Ref.~\cite{Amaral2014} framed in a theory-independent manner, as it determines the largest set allowed by the EP for the experiment, $\mathrm{Y(G)} = \mathrm{abl} \: \mathrm{X(\bar{G})}$, for any assumed correlation set describing the complementary experiment, $\mathrm{X(\bar{G})}$, ranging from the classical to the $E_1$ correlations sets.

We now present our main result, which follows from allowing the complementary experiment to be described by some post-quantum behavior. Interestingly, in this case, the EP leads to an unexpected and somewhat drastic consequence for the description of the experiment itself.

\begin{result}\label{main_res}
Assuming that the EP holds and that post-quantum behaviors are accessible in at least one experiment, it follows that certain genuinely quantum behaviors become forbidden in a related yet completely independent experiment. More precisely, if a post-quantum behaviour is accessible in an experiment $\mathrm{\bar{G}}$, then EP implies that some quantum behavior will be missed in the experiment $\mathrm{G}$.
\begin{proof}
Let $U(\bar{G})$ be the set of behaviors for the experiment $\bar{G}$, where there exist $w^{\prime} \in U(\bar{G})$ corresponding to a post-quantum behaviour; \textit{i.e.}, $w^{\prime} \in \mathrm{E_{1}(\bar{G})}\setminus \mathrm{Q(\bar{G})}$. From Proposition \ref{prop1}, the set of correlations singled out by the EP to describe the experiment is given by $\mathrm{Y(G)} = \mathrm{abl \:}U(\bar{G})$. More explicitly,
\begin{multline}\label{defY}
\mathrm{Y(G)} = \Big\{p \ge 0 | \sum_i p_i p_{i}^{\prime} \le 1 \: \forall \: p^{\prime} \in U(\bar{G}) \Big\}.    
\end{multline}
Using the self-duality property of QT, we have
\begin{multline}
\mathrm{Q(\bar{G})} = \Big\{p^{\prime} \ge 0 | \sum_i p_{i}^{\prime} p_i \le 1 \:\forall\: p \in \mathrm{Q(G)}\Big\};
\end{multline}
thus, for any $\tilde{p}\in \mathrm{E_{1}(\bar{G}) \setminus Q(\bar{G})}$, there exists $p \in \mathrm{Q(G)}$ such that
\begin{equation}
\sum_i \tilde{p}_{i} p_i > 1.
\end{equation}
Let $\tilde{p} = w^{\prime}$; then there exists $p \in \mathrm{Q(G)}$ such that
\begin{equation}
\sum_i p_i w_{i}^{\prime} > 1.
\end{equation}
Any such $p$ is excluded from the definition of $\mathrm{Y(G)}$ in Eq.~\eqref{defY} and, as a matter of fact, $\mathrm{Y(G)} = \mathrm{abl \: U(\bar{G})} \nsupseteq \mathrm{Q(G)}$. It is even possible to show that all excluded behaviors are post-classical (or, if you will, genuinely quantum), that is, $p \in \mathrm{Q(G) \setminus NC(G)}$. Indeed, from the duality between classical and $E_1$ theory used in corollary \ref{cor1}, we have
\begin{multline}
\mathrm{NC(G)} = \mathrm{abl \: E_{1}(\bar{G})} = \Big\{ p \ge 0 | \sum_i p_i p_{i}^{\prime} \le 1 \\ \:\forall\: p^{\prime} \in \mathrm{E_{1}(\bar{G})}\Big\}. 
\end{multline}
As a consequence, for all post-quantum $w^{\prime} \in U(\bar{G})$, any $p \in \mathrm{NC(G)}$ satisfies $\sum_i p_i w_i \le 1$.
\end{proof}
\end{result}

Let’s now explore the implications of our main result. Under the assumptions that the EP holds and some post-quantum correlation is allowed for at least one complementary experiment, our findings ensure that some genuinely quantum behaviors become prohibited. However, quantum theory has been extensively tested in various experimental contexts, and its predictions have been found to agree with observation across an extraordinarily wide range of scenarios \cite{bouwmeester_experimental_1997, razavy_quantum_2003, hanneke_new_2008, juffmann_real-time_2012, Giustina2015, Hensen2015,Shalm2015}. While certain quantum behaviors may be technically challenging to realize \cite{nadlinger_experimental_2022, zhao_loophole-free_2024, bernstein_quantum_1997, watrous2008quantumcomputationalcomplexity},  there is no fundamental reason to believe that any of them is impossible to achieve in Nature. In this way, it seems adequate to assume that every quantum behavior is, in principle, realizable in Nature. Granted, this assumption cannot be empirically verified, as it would require certifying an infinite number of quantum correlations. Nevertheless, it has been explicitly adopted in some works \cite{dirac_principles_2010}, based on the premise that every quantum state and measurement predicted by quantum theory can be realized in Nature—a similar assumption, known as the \textit{no-restriction hypothesis}, is also commonly adopted in the context of GPTs \cite{Chiribella2010, Chiribella2011}. Moreover, we argue that an analogous assumption is implicitly made in many other works \cite{braunstein_wringing_1990, BBDH1997PRL, Cerf_GHZParadoxes_2002, acin_randomness_2012, araujo_all_2013, ZhaoQ2023, vieira2024test}. For instance, it underlies studies in which families of states and measurements—sometimes highly intricate—are assumed, theoretical results are derived from them, and these results are then regarded as valid despite the lack of direct empirical verification. To summarise, since our main result contradicts a premise we firmly uphold (that every quantum behavior is, in principle, realizable in Nature), it is plausible to argue that one of the other two assumptions (that the EP holds and that some experiment might be described by a post-quantum behavior) must be false.

Given that the EP is our chosen physical principle, we will first assume that it is indeed respected in Nature—a view supported by experimental evidence \cite{nawareg2013bounding}. Consequently, we conclude that the EP imposes a fundamental limitation, ruling out the existence of correlation sets that extend beyond the quantum one. This conclusion generalizes ATC's second result in \cite{Amaral2014} to arbitrary experiments, beyond the specific case of self-complementary graphs. At this point, a remark is in order. Our main conclusion is that, given our assumptions, post-quantum correlations are ruled out. While this result primarily concerns correlation sets, it also has indirect implications for post-quantum theories. By a theory, we mean any set of rules that for any given graph $\mathrm{G}$ associates a set of behaviors $\mathrm{U(G)}$, for instance GPTs within the CSW framework. Such a theory will be post-quantum if there exists at least one graph $\mathrm{G}$, such that $\mathrm{U(G)} \nsubseteq  \mathrm{Q(G)}$. Since our results exclude the realization of any such correlation set, it is possible to argue that they also exclude any theory capable of producing it.

Continuing with the assumption that no quantum behavior is fundamentally impossible to realize in Nature, an alternative perspective for those who do not view QT as the ultimate theory of Nature is that the EP does not hold universally. In other words, if a Post-Quantum behavior were ever observed in an experiment, our result would ensure a breakdown of the Exclusivity Principle. However, note that discussing the EP in these post-quantum theories is meaningful only if a notion of sharpness for measurements is assumed within them. This notion of sharpness in Generalized Probabilistic Theories, along with the EP and their relationship, is discussed in \cite{chiribella_measurement_2014, gonda_almost_2018}.

Before concluding this section, we make a brief technical remark. In the proof of our main result, nothing was assumed about the exact form of the correlation set $U(\bar{G})$, other than the existence of some post-quantum behavior in it. Nonetheless, based on the previous discussion, it is only natural to argue that the most adequate form for us to assume for said correlation set is actually $U(\bar{G}) \equiv \mathrm{Q(\bar{G})} \cup W(\bar{G})$, where each $w^{\prime} \in W(\bar{G})$ corresponds to a post-quantum behaviour; \textit{i.e.}, $w^{\prime} \in \mathrm{E_{1}(\bar{G})}\setminus \mathrm{Q(\bar{G})}$. That is, it extends beyond quantum correlations while still containing all of them. In fact, such a form of correlation set seems to be physically interesting, since it is consistent with current empirical data (we are able to obtain quantum bounds of NC inequalities), but might potentially lead to new physics by also allowing the possibility of violating said quantum bounds for some NC inequality. In this particular case, it is straightforward to see Proposition \ref{prop1} yields 
\begin{equation}
\mathrm{Y(G)} = \mathrm{Q(G)} \cap \Big\{ p \ge 0 \:|\: \sum_i p_i w_{i}^{\prime} \le 1 \:\forall\: w^{\prime} \in W(\bar{G})\Big\},
\end{equation}
and it follows immediately that $\mathrm{Y(G)} \subseteq \mathrm{Q(G)}$ (that is, the EP constrains the correlation set of $G$ to be at most quantum). But, as shown before, the existence of post-quantum behaviors in the complementary experiment implies that some genuinely quantum behaviors of the experiment become forbidden, and thus $\mathrm{Y(G)} = \mathrm{abl \: U(\bar{G})} \subset \mathrm{Q(G)}$. In words, any post-quantum correlation set (containing the quantum) assumed for the complementary experiment leads to a correlation set strictly smaller than quantum for the experiment, singled out by the EP. To close this comment, note that even though some genuinely quantum behaviors of the experiment become forbidden, this is not the case for all of them; that is, $\mathrm{Y(G)} = \mathrm{abl \: U(\bar{G})}$, is, in general, a post-classical correlation set, containing $\mathrm{NC(G)}$ as a proper subset. In fact, recalling Corollary \ref{cor1}, $\mathrm{abl \: U(\bar{G})}$ is purely classical only if we choose $\mathrm{U(\bar{G}) = Q(\bar{G})} \cup W(\bar{G}) = \mathrm{E_1(\bar{G})}$.

\section{Conclusion}

In this work, we extended previous results on the quantum correlations problem within the CSW framework. 
Specifically, we examined a type of well-known composite correlation experiment,  Yan's construction.
We identified that the largest correlation set allowed by the Exclusivity Principle (EP) to describe an experiment, given a correlation set for the complementary experiment, aligns with the mathematical structure of anti-blocking sets.
This insight enabled us to determine, for any assumed correlation set of any complementary experiment, the corresponding correlation set selected by the EP for the original experiment (Proposition \ref{prop1}). 
We then explored the implications of this result for classical and $E_1$ sets (Corollary \ref{cor1}), and more importantly, for quantum correlations (Remark \ref{cor2}), where we recovered a previously established result from the literature.

Building on Proposition \ref{prop1} we derived our main result, where we demonstrated that if Nature is at least quantum, then the EP forbids post-quantum behaviors in Nature. 
In other words, by assuming that every quantum behavior is, in principle, realizable in Nature and that the EP holds, we derive a fundamental limitation: the EP excludes the possibility of theories extending beyond Quantum Theory, generalizing previous results to arbitrary experiments. 
Alternatively, if Post-Quantum behaviors are observed, this would indicate that EP is not a valid universal principle. Our work also serves as a bridge between the findings of ATC's 2014 work \cite{Amaral2014} (along with earlier works that follow the same logical structure), by generalizing some of its results, and Cabello's 2019 work \cite{Cabello2019}, by reaching a similar conclusion for the quantum correlations problem. While both approaches highlight the role of the EP, our approach reaches this result through Yan’s more streamlined construction, offering a simpler framework.

It would be interesting to compare our approach with other frameworks that aim to recover quantum theory from general principles. In particular, our results invite a comparison with axiomatic reconstructions developed within the GPT framework, especially in the context of the CSW approach \cite{chiribella_measurement_2014, gonda_almost_2018}. Of special interest is the role of the No-Restriction Hypothesis (NRH) \cite{Chiribella2010, Chiribella2011}, which appears to bear some unveiled relation to the Exclusivity Principle.

To conclude this paper, let us mention a rather interesting question that aligns well with the general ideas presented here. An intriguing observation follows from the self-duality property $\mathrm{Q(\bar{G})} = \mathrm{abl \: Q(G)}$, which QT satisfies. In essence, the theory assumed for the complementary experiment—quantum—is precisely the same as the theory singled out by the EP for the original experiment (see Remark's \ref{cor2} proof). In this context, a natural question arises: are there other GPTs within the CSW framework whose associated correlation sets $U(.)$ satisfy the self-duality relation $\mathrm{U(\bar{G})} = \mathrm{abl \: U(G)}$? If so, Proposition \ref{prop1} would yield conclusions analogous to those derived in the quantum case. If not, QT\footnote{More precisely, the equivalence class of all theories that within the framework of CSW generate the same set of correlations as quantum theory for every graph G.} is uniquely characterized by this self-duality property. Our main result contributes to addressing this issue by showing that any theory satisfying the self-duality relation can neither strictly contain QT nor be strictly contained by it.

\begin{acknowledgments}
We thank Bárbara Amaral, Adán Cabello and Rafael Rabelo for enlightening discussions. This study was financed in part by the Brazilian National Council for Scientific and Technological Development (CNPq) and in part by the Coordenação de Aperfeiçoamento de Pessoal de Nível Superior - Brasil (CAPES) - Finance Code 001. This work was also supported by CNPq under grants no. 311314/2023-6 and 445328/2024-0, and it is part of the Brazilian National Institute for Science and Technology in Quantum Information (INCT-QI). \\
\end{acknowledgments} 

\bibliography{citations}
\end{document}